\documentclass[aps,twocolumn,amsmath,letterpaper]{revtex4}
\usepackage{amssymb}
\usepackage{graphicx}
\usepackage{array}
\usepackage{hhline}
\usepackage{longtable}

\begin{document}

    \title{Overfrustrated and Underfrustrated Spin Glasses in $\textbf{d}=3$ and $2$:

    Evolution of Phase Diagrams and Chaos Including Spin-Glass Order in $\textbf{d}=2$}

    \author{Efe Ilker$^{1}$ and A. Nihat Berker$^{1,2}$}
 \affiliation{$^1$Faculty of Engineering and Natural Sciences, Sabanc\i~University, Tuzla 34956, Istanbul, Turkey,}
    \affiliation{$^2$Department of Physics, Massachusetts Institute of Technology, Cambridge, Massachusetts 02139, U.S.A.}

\begin{abstract}

In spin-glass systems, frustration can be adjusted continuously and
considerably, without changing the antiferromagnetic bond
probability $p$, by using locally correlated quenched randomness, as
we demonstrate here on hypercubic lattices and hierarchical
lattices. Such overfrustrated and underfrustrated Ising systems on
hierarchical lattices in $d=3$ and 2 are studied. With the removal
of just 51\% of frustration, a spin-glass phase occurs in $d=2$.
With the addition of just 33\% frustration, the spin-glass phase
disappears in $d=3$. Sequences of 18 different phase diagrams for
different levels of frustration are calculated in both dimensions.
In general, frustration lowers the spin-glass ordering temperature.
At low temperatures, increased frustration favors the spin-glass
phase (before it disappears) over the ferromagnetic phase and
symmetrically the antiferromagnetic phase. When any amount,
including infinitesimal, frustration is introduced, the chaotic
rescaling of local interactions occurs in the spin-glass phase.
Chaos increases with increasing frustration, as seen from the
increased positive value of the calculated Lyapunov exponent
$\lambda$, starting from $\lambda =0$ when frustration is absent.
The calculated runaway exponent $y_R$ of the renormalization-group
flows decreases with increasing frustration to $y_R=0$ when the
spin-glass phase disappears. From our calculations of entropy and
specific heat curves in $d=3$, it is seen that frustration lowers in
temperature the onset of both long- and short-range order in
spin-glass phases, but is more effective on the former. From
calculations of the entropy as a function of antiferromagnetic bond
concentration $p$, it is seen that the ground-state and
low-temperature entropy already mostly sets in within the
ferromagnetic and antiferromagnetic phases, before the spin-glass
phase is reached.

PACS numbers: 75.10.Nr, 05.10.Cc, 64.60.De, 75.50.Lk



\end{abstract}

    \maketitle
    \def\s{\rule{0in}{0.28in}}
    \setlength{\LTcapwidth}{\columnwidth}

\begin{figure}[]
\centering
\includegraphics[scale=1]{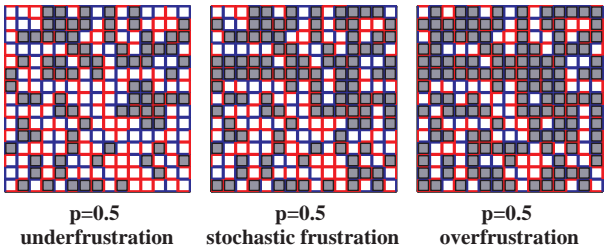}
\caption{(Color online) Randomly distributed ferromagnetic (blue)
and antiferromagnetic (red) interactions on a square plane.  In all
three cases, the antiferromagnetic bond concentration is $p=0.5$.
The frustrated squares are shaded.  In the case at the center, the
bonds were distributed in an uncorrelated fashion, leading to the
frustration of half of the squares (stochastic frustration).  In the
case at the left, 25\% of the frustration was randomly removed
without changing $p=0.5$ (underfrustration).  In the case at the
right, 25\% frustration was randomly added without changing $p=0.5$
(overfrustration). Frustration can thus be set between zero and
complete frustration. It is clear that frustration can thus be
adjusted in all hypercubic lattices.}
\end{figure}

\section{Introduction}

The occurrence of spin-glass long-range order \cite{NishimoriBook},
ground-state entropy \cite{BerkerKadanoff,BerkerKadanoffCo}, and
chaotic rescaling behavior \cite{McKayChaos,McKayChaos2} has long
been discussed in spin-glass systems, with reference to spatial
dimensionality $d$, interaction randomness and frustration
\cite{Toulouse}, accepted as inherent to spin-glass systems and
spin-glass order. In Ising models with randomly distributed
nearest-neighbor ferromagnetic and antiferromagnetic interactions on
hypercubic lattices, it has been shown that a spin-glass phase does
not occur in $d=2$ and does occur in $d=3$.\cite{Binder} In these
hypercubic systems, frustration occurs in elementary squares with an
odd number of antiferromagnetic interactions.  Thus, with
interactions randomly distributed with no correlation, maximally 50
\% of the elementary squares can be frustrated. This fraction
increases from zero as the concentration of frozen antiferromagnetic
bonds $p$ is increased from zero and reaches its maximal value of 50
\% at $p=0.5$.

The basis of the current study is the realization that, for any
value of the antiferromagnetic bond concentration $0<p<1$, the
fraction of frustrated squares can be varied considerably. For
example, for the square lattice, for $0.25 \leq p \leq 0.75$, the
fraction of frustrated squares can be made to vary to any value
between 0 and 1 inclusive, by the locally correlated occurrence
quenched random bonds. For $p \leq 0.25$, the fraction of frustrated
squares can similarly be made to vary between 0 and $4p$.  For $0.75
\leq p$, the fraction of frustrated squares can be made to vary
between 0 and $4(1-p)$. (Thus, frustration reaches 0 with no
variation as $p$ approaches 0 or 1.) Examples are shown in Fig. 1
for $p=0.5$. Thus, when the fraction of frustrated squares is zero,
we have a so-called Mattis spin glass \cite{Mattis}. At the other
extreme, we have a fully frustrated system
\cite{Blankschtein,Bernardi,Murtazaev,Diep1,Diep2}.  All frustration
values in between can be obtained, by randomly removing or adding
local frustration without changing the antiferromagnetic bond
concentration $p$ (Fig. 1).

In this study, we have implemented an exact renormalization-group
study for Ising spin-glass models on the hierarchical lattices, with
$d=3$ and $d=2$, respectively shown in Figs. 2(b) and 3(b), for
arbitrary overfrustration or underfrustration implemented by locally
correlated quenched randomness. We have calculated 18 complete phase
diagrams, each for a different frustration level, in temperature and
antiferromagnetic bond probability $p$. We find that the increase of
frustration disfavors the spin-glass phase (while at low
temperatures favoring the spin-glass phase at the expense of the
ferromagnetic phase and, symmetrically, antiferromagnetic phase.)
Both in $d=3$ and $d=2$, the spin-glass phase disappears at zero
temperature when a certain level of frustration is reached. However,
this disappearance of the spin-glass phase happens in different
regimes in $d=3$ and $d=2$: For $d=3$, it occurs in overfrustration,
so that at stochastic frustration (no correlation in randomness) a
spin-glass phase occurs.  For $d=2$, it already occurs in
underfrustration, so that at stochastic frustration a spin-glass
phase does not occur. However, with frustration only partially
removed, we find that a spin-glass phase certainly does occur in
$d=2$.

The chaotic rescaling
\cite{McKayChaos,McKayChaos2,BerkerMcKay,Hartford,Frzakala1,Frzakala2,Sasaki,Lukic,Ledoussal,Rizzo,Katzgraber,Yoshino,
Aspelmeier1,Aspelmeier2,Mora,Aral,Stone,Jorg,Obuchi,Ilker,Roma,Chen,Fernandez,deLima}
of the interactions within the spin-glass phase occurs as soon as
frustration is increased from zero, both in $d=3$ and $d=2$. We have
calculated the Lyapunov exponent $\lambda$ \cite{Collet,Hilborn} of
the renormalization-group trajectory of the interaction at a given
location, when the system is in the spin-glass phase. When
frustration is increased from zero, the Lyapunov exponent $\lambda$
increases from zero, both in $d=3$ and $d=2$. This behavior is of
course consistent with the chaotic renormalization-group
trajectories. Different values of the positive Lyapunov exponents
characterize different spin-glass phases. It is found here that the
value of the Lyapunov exponent continuously varies with the level of
frustration and is different for each dimensionality $d$. The
Lyapunov exponent does not depend on antiferromagnetic bond
concentration $p$ or temperature.

Our calculations with varying frustration also yield information on
long- and short-range ordering, and entropy.  The increase in
frustration lowers both the onset temperature of long-range order
and the characteristic temperature of short-range order, but affects
long-range order much more drastically, thus interchanging the two
temperatures and eventually eliminating long-range spin-glass order.
For $d=3$, for low frustration, the specific heat peak occurs inside
the spin-glass phase, indicating that considerable short-range
disorder persists into the higher temperatures of the spin-glass
phase.  In these cases, as temperature is lowered, spin-glass
long-range order onsets before the system is predominantly
short-range ordered.  As frustration is increased, both ordering
temperatures are lowered, but differently, so that they interchange
before stochastic frustration is reached. Thus, for overfrustration,
stochastic frustration, and higher frustration values of
underfrustration, the specific heat peak occurs outside the
spin-glass phase, indicating that as temperature is lowered,
short-range order sets before long-range order (which reaches zero
temperature in overfrustration). Zero-temperature or low-temperature
entropy is a distinctive character of systems with frustration.
Frustration is introduced into the system, by increasing from zero
the antiferromagnetic bond concentration $p$. It is seen that
frustration favors the spin-glass phase over the ferromagnetic
phase.  However, it is also seen that, in all cases that frustration
is introduced, the major portion of the entropy is created with the
ferromagnetic phase as opposed to the spin-glass phase.

\section{Overfrustrated and underfrustrated spin-glass systems on hypercubic lattices and hierarchical lattices}

\subsection{Stochastic Frustration, Overfrustration, and Underfrustration on Hypercubic Lattices}

The Ising spin-glass model is defined by the Hamiltonian
\begin{equation}
-\beta \mathcal{H}=\sum_{\langle ij \rangle} J_{ij} s_i s_j
\end{equation}
where $\beta=1/kT$, at each site $i$ of a lattice the spin $s_i =
\pm 1$, and $\langle ij \rangle$ denotes that the sum runs over all
nearest-neighbor pairs of sites. The bond strengths $J_{ij}$ are
$+J>0$ (ferromagnetic) with probability $1-p$ and $-J$
(antiferromagnetic) with probability $p$. On hypercubic lattices, in
any elementary square with an odd number number of antiferromagnetic
bonds, all bonds cannot be simultaneously satisfied, meaning that
there is frustration.\cite{Toulouse}  When the antiferromagnetic
bonds are randomly distributed with probability $p$ across the
lattice, a fraction
\begin{equation}
4p(1-p)^3 + 4p^3(1-p) = 4(p- 3p^2 + 4p^3 -2p^4)
\end{equation}
of the elementary squares is frustrated. This system with
uncorrelated quenched randomness is the usually studied spin-glass
system and we shall refer to it as a \textbf{stochastically
frustrated} system.  On the other hand, by changing the signs of
individual bonds $J_{ij} \rightarrow -J_{ij}$ at randomly chosen
localities, with the rule that, for every
ferromagnetic-to-antiferromagnetic local change, an
antiferromagnetic-to-ferromagnetic local change is done, frustration
can be continuously increased or decreased from the value in Eq.(2),
without changing the antiferromagnetic bond concentration $p$. We
call the systems in which frustration is thus increased or decreased
from stochastic frustration, respectively, \textbf{overfrustrated}
or \textbf{underfrustrated} systems. Examples of overfrustration,
stochastic frustration, and underfrustration are given in Fig. 1.

\begin{figure}[h!]
\centering
\includegraphics[scale=1]{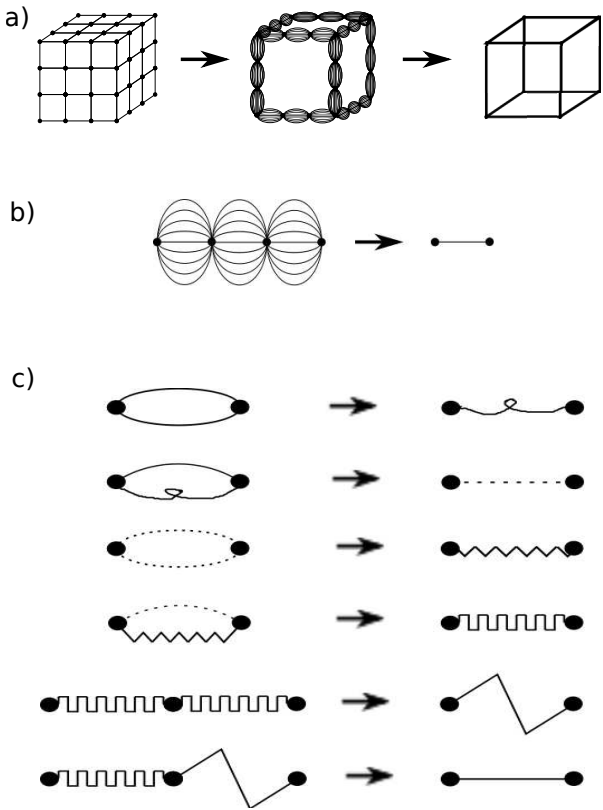}
\caption{(a) Migdal-Kadanoff approximate renormalization-group
transformation for the $d=3$ cubic lattice with the length-rescaling
factor of $b=3$. Bond-moving is followed by decimation. (b) Exact
renormalization-group transformation for the equivalent $d=3$
hierarchical lattice with the length-rescaling factor of $b=3$. (c)
Pairwise applications of the quenched probability convolution of
Eq.(5), leading to the exact transformation in (b) and, numerically
equivalently, to the approximate transformation in (a).}
\end{figure}

\begin{figure}[h!]
\centering
\includegraphics[scale=1]{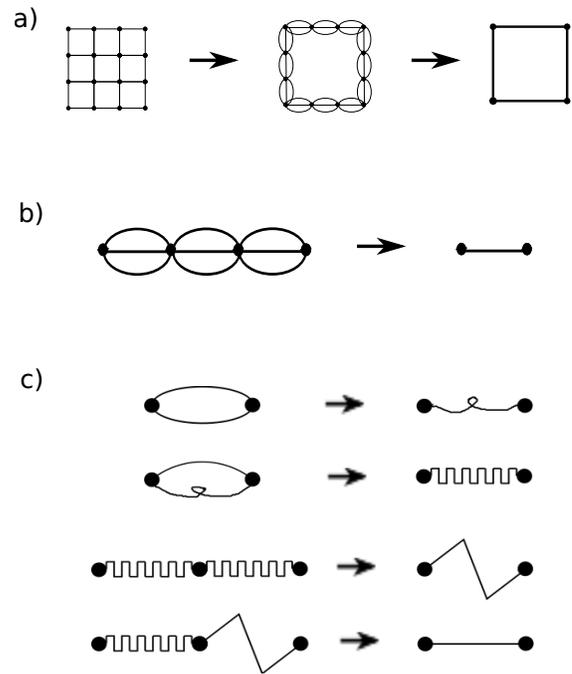}
\caption{(a) Migdal-Kadanoff approximate renormalization-group
transformation for the $d=2$ square lattice with the
length-rescaling factor of $b=3$. Bond-moving is followed by
decimation. (b) Exact renormalization-group transformation for the
equivalent $d=2$ hierarchical lattice with the length-rescaling
factor of $b=3$. (c) Pairwise applications of the quenched
probability convolution of Eq.(5), leading to the exact
transformation in (b) and, numerically equivalently, to the
approximate transformation in (a).}
\end{figure}

\subsection{Renormalization-Group Transformation, Quenched Probability Convolutions by Histograms and Cohorts}

The usual, stochastically frustrated spin-glass systems on
hypercubic lattices are readily solved by a renormalization-group
method that is approximate on the hypercubic lattice
\cite{Migdal,Kadanoff} and simultaneously exact on the hierarchical
lattice \cite{BerkerOstlund,Kaufman1,Kaufman2,McKay,Hinczewski1}.
Under rescaling, the form of the interaction as given in Eq.(1) is
conserved. The renormalization-group transformation, for spatial
dimension $d$ and length-rescaling factor $b = 3$ (necessary for
treating the ferromagnetic and antiferromagnetic correlations on
equal footing), is achieved (Figs. 2(a) and 3(a)) by a sequence of
bond moving
\begin{equation}
J_{ij}^{(bm)} = \sum_{<kl>}^{b^{d-1}} J_{kl}
\end{equation}
and decimation
\begin{equation}
e^{J_{im}^{(dec)}s_i s_m + G_{im}}=\sum_{s_j,s_k} e^{J_{ij} s_i
s_j+J_{jk} s_j s_k +J_{km} s_k s_m},
\end{equation}
where the additive constants $G_{ij}$ are unavoidably generated.

\begin{figure}[]
\centering
\includegraphics[scale=0.9]{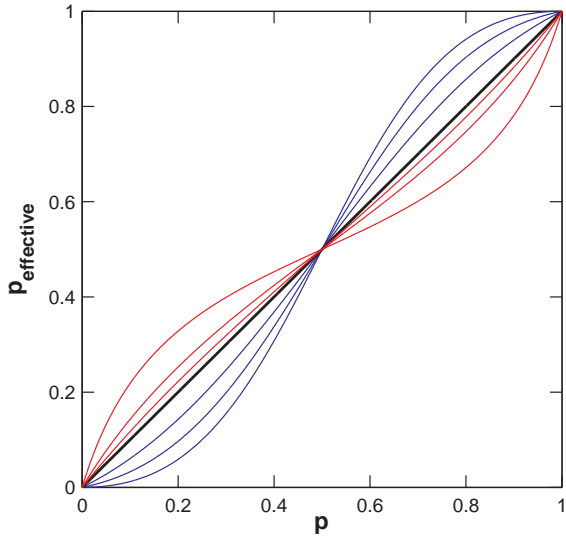}
\caption{(Color online) $p_{effective}$ versus $p$ for the range of
underfrustration and overfrustration used in our study (Eq.(6)). The
curves are, consecutively from the lower right, for $f=0,0.2,0.5;
f=1=g$ (thicker line); $g=0.8,0.6,0.3$.}
\end{figure}

The starting bimodal quenched probability distribution of the
interactions, characterized by $p$ and described above, is not
conserved under rescaling. The renormalized quenched probability
distribution of the interactions is obtained by the convolution
\cite{Andelman}
\begin{equation}
\label{eq:5} P'(J'_{i'j'}) = \int{\left[\prod_{ij}^{i'j'}dJ_{ij}
P(J_{ij})\right]} \delta(J'_{i'j'}-R(\left\{J_{ij}\right\})),
\end{equation}
where the primes denote the renormalized system and
$R(\left\{J_{ij}\right\})$ represents the bond moving and decimation
given in Eqs.(3) and (4). For numerical practicality, the bond
moving and decimation of Eqs.(3) and (4) are achieved by a sequence
of pairwise combination of interactions, as shown for $d=3$ and
$d=2$ respectively in Figs. 2(c) and 3(c), each pairwise combination
leading to an intermediate probability distribution resulting from a
pairwise convolution as in Eq.(5).

We implement this procedure numerically in two calculationally
equivalent ways:  (1) The quenched probability distribution is
represented by
histograms.\cite{Migliorini,Hinczewski,Guven,Gulpinar}  A total
number of between 500 to 2,500 histograms, depending on the needed
accuracy, is used here. This total number is distributed between
ferromagnetic $J>0$ and antiferromagnetic $J<0$ interactions
according to the total probabilities for each case. (2) By
generating a cohort of 20,000 interactions \cite{Ilker} that
embodies the quenched probability distribution. At each pairwise
convolution as in Eq.(5), 20,000 randomly chosen pairs are matched
by Eq.(3) or (4), and a new set of 20,000 is produced. The numerical
convergence of the histogram and cohort implementations are
determined, respectively, by the numbers of histograms and cohort
members.  At numerical convergence, the results of the two
implementations match. The histogram method is faster and is used to
calculate phase diagrams, thermodynamic properties, and asymptotic
fixed distributions.  The cohort method is needed for studying the
repeated rescaling behavior of the interaction at a specific
location on the lattice and is used to calculate chaotic
trajectories, chaotic bands, and Lyapunov exponents.\cite{Ilker}

\subsection{Stochastic Frustration, Overfrustration, and Underfrustration on Hierarchical Lattices}

Hierarchical models are models which are exactly soluble by
renormalization-group
theory.\cite{BerkerOstlund,Kaufman1,Kaufman2,McKay,Hinczewski1}
Hierarchical lattices have therefore been used to study a variety of
spin-glass and other statistical mechanics
problems.\cite{Gingras2,Migliorini,Gingras1,Hinczewski,Guven,Gulpinar,Shrock,Herrmann1,Hermann2,Garel,Hartmann,Fortin,Wu}
Hierarchical models can be constructed \cite{BerkerOstlund} that
have identical renormalization-group recursion relations with the
approximate treatment of models on hypercubic and other Euclidian
lattices.  Thus, Figs. 2(b) and 3(b) respectively give the
hierarchical models, used in our study, that have the same recursion
relations as the Migdal-Kadanoff approximation
\cite{Migdal,Kadanoff} for the hypercubic lattice in $d=3$ (cubic
lattice) and $d=2$ (square lattice).

Overfrustration or underfrustration is readily introduced into
hierarchical lattices by randomly changing local interactions or
groups of local interactions, while conserving $p$.  This
overfrustration or underfrustration affects the pairwise bond-moving
step of the renormalization-group solution. In the case of
overfrustration, when two bonds are matched for bond-moving, bonds
of the same sign are accepted with a probability $g$, $0\leqslant g
< 1$. Clearly, when $g=1$, we have not altered the occurrence of
frustration.  But, for a value of $g$ in the range $0\leqslant g <
1$, we have removed a fraction $1-g$ of the unfrustrated
occurrences.

Similarly, in the case of underfrustration, when two bonds are
matched for bond-moving, bonds of the opposite sign are accepted
with a probability $f$, $0\leqslant f < 1$.  Again, when $f=1$, we
have not altered the occurrence of frustration.  But, for a value of
$f$ in the range $0\leqslant f < 1$, we have removed a fraction
$1-f$ of the frustrated occurrences.

We have thus defined the degree of frustration on the hierarchical
models. Accordingly, full frustration, stochastic frustration, and
zero frustration respectively correspond to $g=0$, $g=1=f$, $f=0$.
Our implementation of underfrustration and overfrustration via the
factors $f$ and $g$ does affect, on the hierarchical lattice, the
effective value of the antiferromagnetic bond probability $p$ as
\begin{equation}
\begin{split}
p_{effective} & = \frac{p-(1-f)p(1-p)}{1-(1-f)2p(1-p)},\\
p_{effective} & =\frac{p-(1-g)p^2}{1-(1-g)(p^2+(1-p)^2)}.
\end{split}
\end{equation}
$p_{effective}$ includes the combined effect of $p$ together with
the local quenched correlation rule controlled by $f$ or $g$.  (The
actual microscopic renormalization-group calculation is of course
done using $p$ with the quenched correlation rule, which completely
defines the model.) Eqs.(6) directly follow from the acceptance
rules given in the previous two paragraphs:  The second terms in the
numerators subtract the probability due to rejection because of a
bond-moving match that is suppressed; the denominator is a
normalization taking into account this rejection probability.  Thus,
$p=0.5$, the center of a would-be spin-glass phase, is not affected.
For other values, $p_{effective}$ stays close to $p$, as seen in
Fig. 4. Just as in the case of underfrustrated and overfrustrated
hypercubic lattices (Fig. 1), underfrustrated and overfrustrated
hierarchical lattices as defined and studied here can be physically
realized. However, our procedure of underfrustrating or
overfrustrating hierarchical lattices is not a direct representation
of underfrustrating or overfrustrating hypercubic lattices.  One
important difference is that, in hierarchical lattices,
underfrustrating or overfrustrating is done at every length scale.
This leaves the underfrustrated or overfrustrated hypercubic
lattices, which can be achieved as we demonstrated, as an
interesting open problem, with our current results only being
suggestive.

\begin{figure*}[]
\centering
\includegraphics*[scale=0.8]{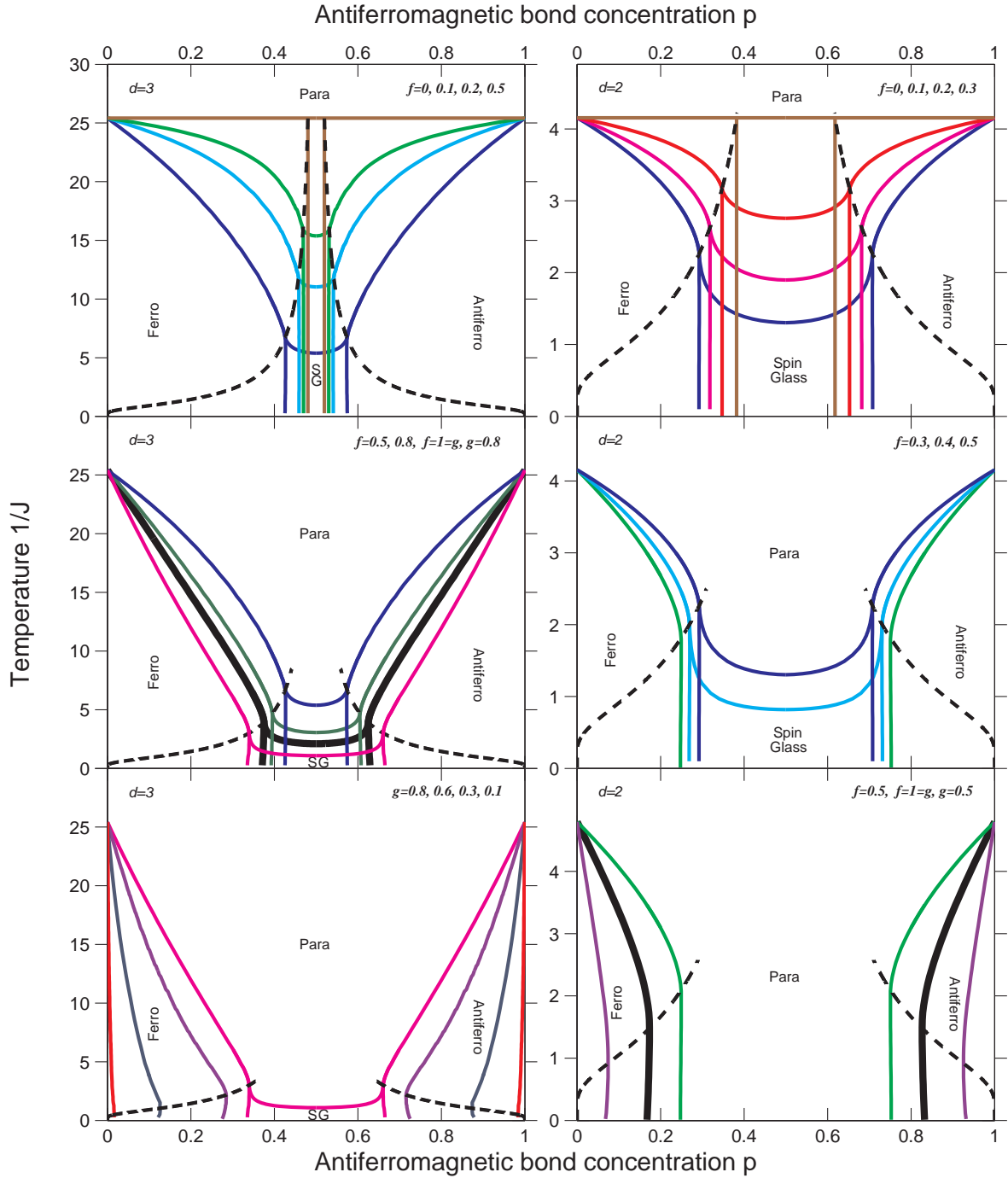}
\caption{(Color online) Calculated phase diagrams of the
overfrustrated, underfrustrated, and stochastically frustrated Ising
spin-glass models on hierarchical lattices. The panels on the left
side are for $d=3$ dimensions. Left top panel: The outermost phase
diagram, consisting of one horizontal and two vertical lines, is for
no frustration, $f=0$.  Starting from this outermost phase diagram,
the three consecutive phase diagrams are for the underfrustrated
cases (where frustration has been removed) of $f=0.1,0.2,0.5$. Left
middle panel: Starting from the outermost phase diagram, the four
consecutive phase diagrams are for the underfrustrated cases of
$f=0.5,0.8$; the stochastic case (where frustration has been neither
removed, nor added) of $f=1=g$, drawn with the thicker lines; and
the overfrustrated case (where frustration has been added) of
$g=0.8$.  Left bottom panel: Starting from the outermost phase
diagram, the four consecutive phase diagrams are for the
overfrustrated cases of $g=0.8,0.6,0.3,0.1$. In the latter three
cases, $g=0.6,0.3,0.1$, no spin-glass phase occurs. Excessive
overfrustration destroys the spin-glass phase. The panels on the
right side are for $d=2$ dimensions. Right top panel: The outermost
phase diagram, consisting of one horizontal and two vertical lines,
is for no frustration, $f=0$.  Starting from this outermost phase
diagram, the three consecutive phase diagrams are for the
underfrustrated cases of $f=0.1,0.2,0.3$. Right middle panel:
Starting from the outermost phase diagram, the three consecutive
phase diagrams are for the underfrustrated cases of $f=0.3,0.4,0.5$.
Right bottom panel: Starting from the outermost phase diagram, the
three consecutive phase diagrams are the underfrustrated case of
$f=0.5$; for the stochastic case of $f=1=g$, drawn with the thicker
lines; and the overfrustrated case of $g=0.5$. In the latter three
cases, $f=0.5,f=1=g, g=0.5$, no spin-glass phase occurs. However, in
the underfrustrated cases of $f=0.1,0.2,0.3,0.4$, a spin-glass phase
occurs in these $d=2$ dimensional systems with locally correlated
randomness. All phase transitions in this figure are second order
and, to the resolution of the figure, all multicritical points
appear on the Nishimori symmetry line, shown with the dashed
curves.}
\end{figure*}

\begin{figure}[]
\centering
\includegraphics[scale=1]{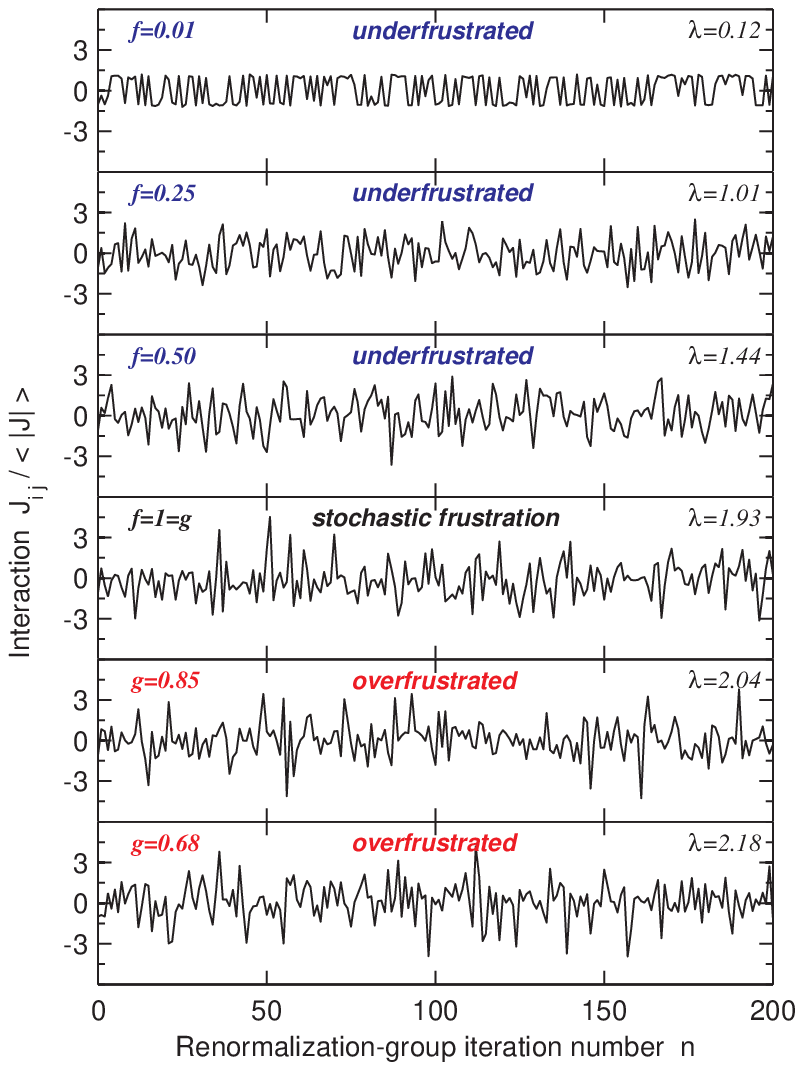}
\caption{(Color online) Interaction at a given position in the
lattice at successive renormalization-group iterations, for $d=3$
systems with different frustrations. In all cases, the
antiferromagnetic bond concentration is $p=0.5$ and the initial
temperature is $1/J=0.2$, inside the spin-glass phase.  For each
frustration amount, a chaotic trajectory of the interaction at a
given position is seen from this figure. The calculated Lyapunov
exponent for each case is given in the upper right corner of each
panel.}
\end{figure}

\begin{figure}[]
\centering
\includegraphics[scale=1]{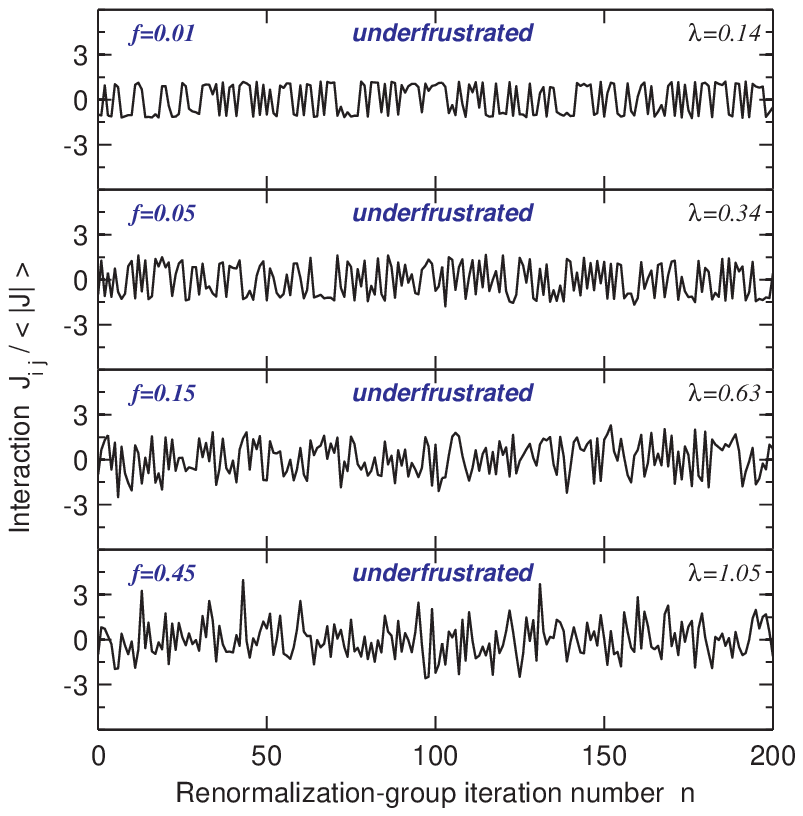}
\caption{(Color online) Interaction at a given position in the
lattice at successive renormalization-group iterations, for $d=2$
systems with different frustrations. In all cases, the
antiferromagnetic bond concentration is $p=0.5$ and the initial
temperature is $1/J=0.2$, inside the spin-glass phase.  For each
frustration amount, a chaotic trajectory of the interaction at a
given position is seen from this figure. The calculated Lyapunov
exponent for each case is given in the upper right corner of each
panel.}
\end{figure}

\begin{figure}[]
\centering
\includegraphics[scale=1]{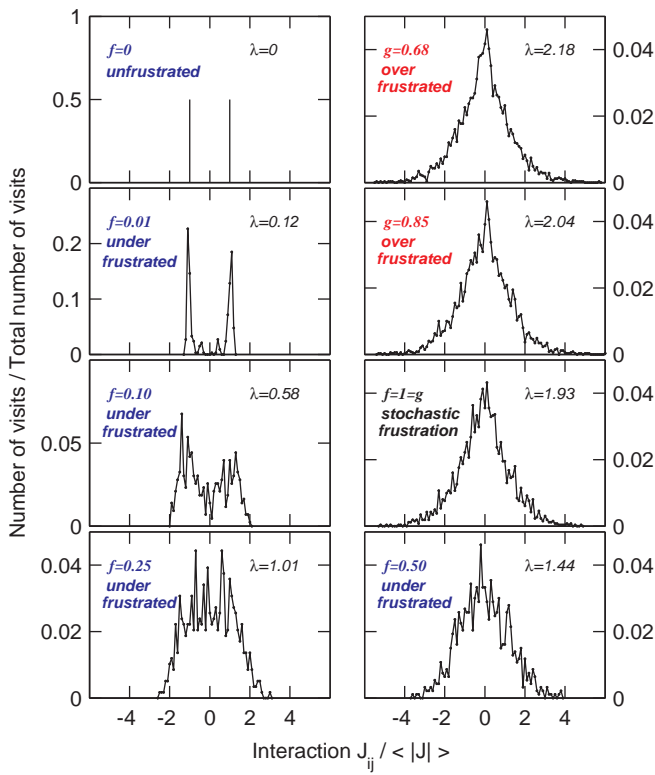}
\caption{(Color online) The chaotic visits of the consecutively
renormalized interactions $J_{ij}$ at a given position of the
system, in the spin-glass phase of overfrustrated, underfrustrated,
and stochastically frustrated Ising models in $d=3$. These
consecutively renormalized interactions at a given position of the
system are shown here as scaled with the average interaction $<|J|>$
across the system, which diverges as $b^{n y_R}$ where $n$ is the
number of renormalization-group iterations and $y_R > 0$ is the
runaway exponent shown in Fig. 10.  The number of visits into each
interval of 0.1 on the horizontal axis have been scaled with the
total number of renormalization-group iterations.  Between 300 and
3,500 renormalization-group iterations have been used for the
different panels.  The distributions of chaotic visits shown in the
panels stabilize as the number of iterations is increased. The
calculated Lyapunov exponent for each case is given in the upper
right corner of each panel.}
\end{figure}

\begin{figure}[]
\centering
\includegraphics[scale=1]{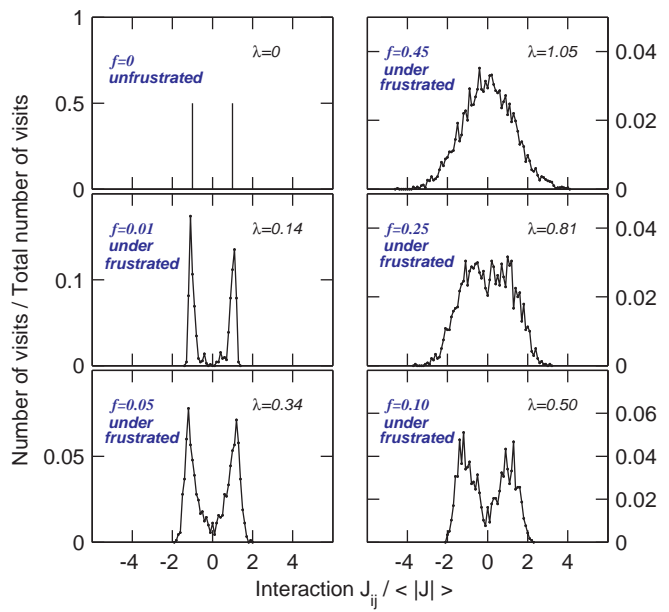}
\caption{(Color online) The chaotic visits of the consecutively
renormalized interactions $J_{ij}$ at a given position of the
system, in the spin-glass phase of underfrustrated Ising models in
$d=2$. These consecutively renormalized interactions at a given
position of the system are shown here as scaled with the average
interaction $<|J|>$ across the system, which diverges as $b^{n y_R}$
where $n$ is the number of renormalization-group iterations and $y_R
> 0$ is the runaway exponent shown in Fig. 10. The number of visits
into each interval of 0.1 on the horizontal axis have been scaled
with the total number of renormalization-group iterations. Between
700 and 5,000 renormalization-group iterations have been used for
the different panels.  The distributions of chaotic visits shown in
the panels stabilize as the number of iterations is increased. The
calculated Lyapunov exponent for each case is given in the upper
right corner of each panel. No spin-glass phase occurs for $f>0.49$,
as seen in Figs. 5 and 10.}
\end{figure}

\begin{figure}[]
\centering
\includegraphics[scale=1]{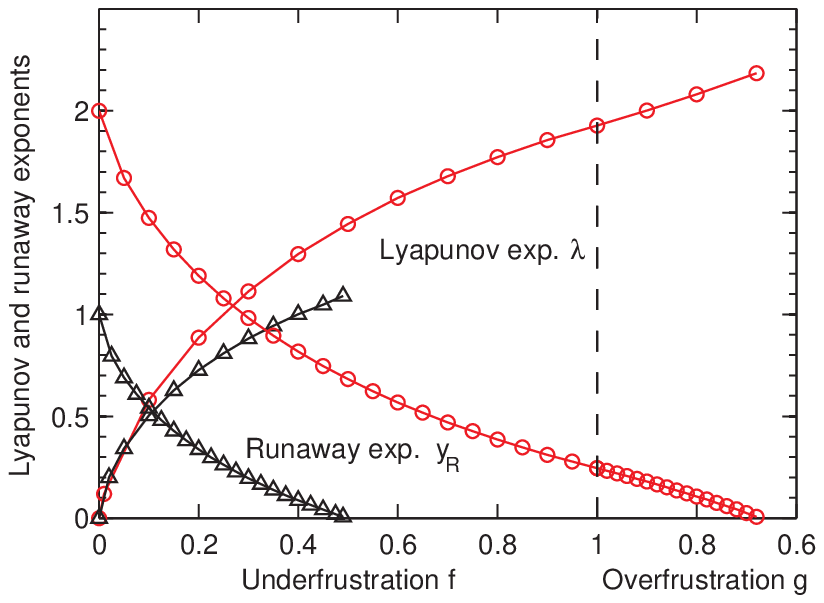}
\caption{(Color online) Lyapunov exponent $\lambda$ and runaway
exponent $y_R$ of the spin-glass phases of overfrustrated,
underfrustrated, and stochastically frustrated Ising models in $d=3$
(upper curves) and $d=2$ (lower curves). The horizontal scale shows,
to the left of the dashed line, the $f$ values of the
underfrustrated cases and, to the right of the dashed line, the $g$
values of the overfrustrated cases. The dashed line marks the
stochastic frustration $(f=1=g)$. As seen in this figure and in
Figs. 8 and 9, as soon as frustration is introduced, $(f>0)$, the
Lyapunov exponent becomes positive and chaotic behavior occurs
inside the spin-glass phase. The average interaction $<|J|>$ across
the system diverges as $b^{n y_R}$ where $n$ is the number of
renormalization-group iterations and $y_R > 0$ is the runaway
exponent.  The Lyapunov exponent $\lambda$ monotonically increases
with frustration from $\lambda=0$ at zero frustration and the
runaway exponent $y_R$ monotonically decreases with frustration from
$y_R=d-1$ at zero frustration.  The spin-glass phase disappears when
$y_R$ reaches zero, for $g=0.67$ in $d=3$ and $f=0.49$ in $d=2$.}
\end{figure}

\subsection{Determination of the Phase Diagrams and Thermodynamic Properties}

The different thermodynamic phases of the model are identified by
the different asymptotic renormalization-group flows of the quenched
probability distributions.  For all renormalization-group flows,
inside the phases and on the phase boundaries, Eq.(5) is iterated
until asymptotic behavior is reached, meaning that we are studying
an effectively infinite hierarchical lattice. The thermodynamic
properties, such as free energy, energy, entropy, and specific heat,
are calculated by summing along entire renormalization-group
trajectories.\cite{BerkerOstlund,McKay,Hinczewski1,Fisher} Thus, we
are able to calculate phase diagrams and thermodynamic properties
for any case of overfrustration or underfrustration.

\section{Calculated phase diagrams for overfrustration and underfrustration in $d=3$ and $d=2$}

Figure 5 shows 18 different calculated phases diagrams, in
temperature $1/J$ and antiferromagnetic bond concentration $p$, for
overfrustrated, stochastically frustrated, underfrustrated Ising
spin-glass models in $d=3$ and $d=2$. Each phase diagram has a
different amount of overfrustration or underfrustration, or is
stochastically frustrated. In general, increased frustration drives
the spin-glass phase to lower temperatures. Thus, the spin-glass
phase disappears at a threshold amount of frustration. This
threshold frustration is dramatically different in $d=3$ and $d=2$,
as explained below. On the other hand, increased frustration favors
the spin-glass phase (before it disappears) over the ferromagnetic
phase and symmetrically the antiferromagnetic phase, at low
temperatures.

The left panels are for $d=3$ dimensions. The outermost phase
diagram, consisting of one horizontal and two vertical lines, is for
no frustration, $f=0$. Starting from this outermost phase diagram,
the consecutive phase diagrams have increasing frustration: They are
for the underfrustrated cases (where frustration has been removed)
of $f=0.1,0.2,0.5,0.8$; the stochastic case (where frustration has
been neither removed, nor added) of $f=1=g$, drawn with the thicker
lines; and the overfrustrated case (where frustration has been
added) of $g=0.8,0.6,0.3,0.1$. In the latter three cases,
$g=0.6,0.3,0.1$, no spin-glass phase occurs. Thus, in $d=3$,
excessive overfrustration destroys the spin-glass phase.

The right panels are for $d=2$ dimensions. Again, the outermost
phase diagram, consisting of one horizontal and two vertical lines,
is for no frustration, $f=0$. Starting from this outermost phase
diagram, the consecutive phase diagrams again have increasing
frustration: They are for the underfrustrated cases of
$f=0.1,0.2,0.3,0.4,0.5$; the stochastic case of $f=1=g$, drawn with
the thicker lines; and the overfrustrated case of $g=0.5$. In the
latter three cases, $f=0.5,f=1=g, g=0.5$, no spin-glass phase
occurs. However, in the underfrustrated cases of
$f=0.1,0.2,0.3,0.4$, a spin-glass phase does occur in these $d=2$
dimensional systems with locally correlated randomness. Thus, when
frustration is increased from zero, the spin-glass phase disappears
while still in the underfrustrated regime. Accordingly, in
ordinarily studied spin-glass systems, which are stochastically
frustrated systems, the spin-glass phase is seen in $d=3$, but not
seen in $d=2$.

The paramagnetic-ferromagnetic-spinglass reentrance for the phase
diagrams with the spin-glass phase and the
paramagnetic-ferromagnetic-paramagnetic (true) reentrance for the
phase diagrams without the spin-glass phase, as temperature is
lowered, is seen here.  Both types of phase diagrams were first
noted with hierarchical models for Ising spin glasses
\cite{Migliorini} and Potts spin glasses \cite{Gingras1}. Phase
diagram reentrance is also seen in experimental spin-glass systems
\cite{Roy} and, most proeminently, in liquid crystal systems where
annealed (as opposed to quenched as in the current study)
frustration plays a role.\cite{Indekeu,Netz,Mazza,ChenS} All phase
transitions in Fig. 5 are second order and, to the resolution of the
figure, the multicritical points appear on the Nishimori symmetry
line, shown with the dashed
curves.\cite{Nishimori1,Nishimori2,Nishimori3,Nishimori4,Nishimori5}

\section{Chaos in the Spin-Glass Phase Triggered by Infinitesimal Frustration}

The local interaction at a given position in the lattice at
successive renormalization-group transformations, in systems with
different frustrations, is given for $d=3$ and 2 respectively in
Figs. 6 and 7. These consecutively renormalized interactions at a
given position of the system are shown here as scaled with the
average interaction $<|J|>$ across the system, which diverges as
$b^{n y_R}$ where $n$ is the number of renormalization-group
iterations and $y_R > 0$ is \textbf{the runaway exponent} shown in
Fig. 10. This divergence indicates \textbf{strong-coupling chaotic
behavior}.\cite{Ilker} In Figs. 6 and 7, it is seen that, for any
amount of frustration, the local interaction at a given position in
the lattice exhibits, under renormalization-group transformations, a
chaotic trajectory.\cite{Hartford}

The cumulative pictures of the chaotic visits of the consecutively
renormalized interactions $J_{ij}$ at a given position of the
system, for a large number of renormalization-group iterations, in
the spin-glass phases for different frustrations, is given for $d=3$
and 2 respectively in Figs. 8 and 9. It has been recently shown
\cite{Ilker} that these distributions over renormalization-group
iterations for a given position in the lattice are completely
equivalent to the distributions of interactions across the lattice
at a given renormalization-group iteration. As seen in Figs. 8 and
9, in the system where frustration is completely removed ($f=0$,
uppermost leftside diagrams), the interaction at a given position
randomly visits positive and negative values, giving the two delta
functions seen in the figures. When frustration is introduced ($f$
is increased from 0), these two delta functions broaden into two
chaotic bands (seen in the figures for $f=0.01$), which merge into a
double-peaked single band (seen for $f=0.10$), which transforms into
a single peak (seen for $f=0.25$). In $d=3$, the single-peaked
chaotic band continues through the stochastic frustration ($f=1=g$)
into a range of overfrustrated systems ($g>0.67$), albeit with
varying Lyapunov exponents $\lambda$, as seen in the insets and in
Fig. 10. In $d=2$, the single-peaked chaotic band continues when
frustration is increased to $f=0.45$ (uppermost rightside diagram),
but no spin-glass phase occurs for $f>0.49$, that is to say in
overfrustration, stochastic frustration, and the higher range of
underfrustration.

The spin-glass phases, being chaotic, can be characterized
\cite{Ilker} by \textbf{the Lyapunov exponent} of general chaotic
behavior \cite{Collet,Hilborn}. The positivity of the Lyapunov
exponent measures the strength of the chaos \cite{Collet,Hilborn}
and was also used in the previous spin-glass study of Ref.[23].  The
calculation of the Lyapunov exponent is applied here to the chaotic
renormalization-group trajectory at any specific position in the
lattice,
\begin{equation}
\lambda = \lim _{n\rightarrow\infty} \frac{1}{n} \sum_{k=0}^{n-1}
\ln \Big|\frac {dx_{k+1}}{dx_k}\Big|
\end{equation}
where $x_k = J_{ij}/<|J|>$ at step $k$ of the renormalization-group
trajectory.  The sum in Eq.(6) is to be taken within the asymptotic
chaotic band.  Thus, we throw out the first 100
renormalization-group iterations to eliminate the points outside of,
but leading to the chaotic band. Subsequently, typically using up to
2,000 renormalization-group iterations in the sum in Eq.(6) assures
the convergence of the Lyapunov exponent value.  The calculated
Lyapunov exponents $\lambda$ and runaway exponents $y_R$ of the
spin-glass phases of overfrustrated, underfrustrated, and
stochastically frustrated Ising models in $d=3$ (upper curves) and
$d=2$ (lower curves) are given in Fig. 10. As seen in this figure
and in Figs. 6-9, as soon as frustration is introduced $(f>0)$, the
Lyapunov exponent becomes positive and chaotic behavior occurs
inside the spin-glass phase.  Upon further increasing frustration,
on the other hand, the spin-glass phase disappears when $y_R$
reaches zero as seen in Fig. 10, for $g=0.67$ in $d=3$ and $f=0.49$
in $d=2$.

\section{Entropy, Short- and Long-Range Order in Overfrustrated and Underfrustrated Spin Glasses}

Information about the relative shift and interchange in short- and
long-range order can be deduced from entropy and specific heat
curves. Short-range order is deduced from a specific heat peak (loss
of entropy) that is away from the phase transition. Long-range order
is deduced from the phase transition given by the
renormalization-group flows.  Thus, the characteristic temperature
of short-range order is the temperature of the specific heat peak.
The characteristic temperature of long-range order is the phase
transition temperature.  The calculated entropy per site $S/kN$ and
specific heat per site $C/kN$ are shown in Fig. 11 as a function of
temperature $1/J$ at fixed antiferromagnetic bond concentration
$p=0.5$, for $d=3$ systems with underfrustration $(f=0.02,0.2,0.5)$,
stochastic frustration $(f=1=g)$, and overfrustration $(g=0.7)$. The
tick mark shows the phase transition point between the spin-glass
phase and the paramagnetic phase for each frustration case. As also
seen in Fig. 5, frustration lowers this transition temperature. For
stochastic frustration ($f=1=g$), the specific heat peak occurs
outside the spin-glass phase, indicating that considerable
short-range ordering occurs at higher temperatures before the onset
of spin-glass long-range order. By contrast, for low frustration
$(f=0.02,0.2)$, the specific heat peak occurs inside the spin-glass
phase, indicating that considerable short-range disorder persists
into the higher temperatures of the spin-glass phase. This
conclusion is also reached from the entropy curves in the upper
panel. The changeover between these two regimes occurs for the
underfrustrated system of $f=0.5$. Overfrustrated systems show
understandably specific heat behavior similar to $f=1$, with
frustration lowering the long-range order temperature and
short-range order setting above this temperature with a specific
heat peak.

The calculated entropy per site $S/kN$ as a function of the
antiferromagnetic bond concentration $p$ at fixed temperature
$1/J=0.5$ is shown in the upper panel of Fig. 12 for $d=3$ systems
with no frustration $(f=0)$, underfrustration $(f=0.5,0.8)$,
stochastic frustration $(f=1=g)$, and overfrustration $(g=0.8)$.
Frustration is thus introduced at different rates in the different
curves in Fig. 12. Here the tick mark shows the phase transition
point between the ferromagnetic phase and the spin-glass phase for
each frustration case. It is seen that frustration favors the
spin-glass phase over the ferromagnetic phase.  It is also seen
that, as soon as frustration is introduced, the major portion of the
entropy is created with the ferromagnetic phase as opposed to the
spin-glass phase. Fig. 12 also shows the calculated derivative of
the entropy per site $(1/kN)(\partial S /k
\partial p)$ as a function of the antiferromagnetic bond
concentration $p$ at fixed temperature $1/J=0.5$, for the stochastic
frustration system $(f=1)$ in $d=3$. The tick mark again marks the
phase transition point between the ferromagnetic phase and the
spin-glass phase.  The peak being inside the ferromagnetic phase
also indicates that short-range disorder sets inside the
ferromagnetic phase.

\begin{figure}[h!]
\centering
\includegraphics[scale=1]{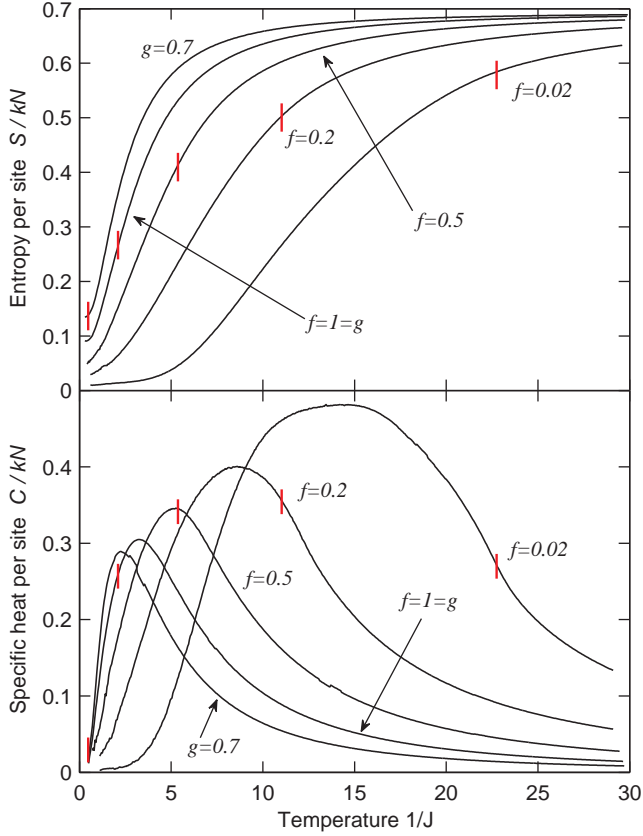}
\caption{(Color online) The calculated entropy per site $S/kN$
(upper panel) and specific heat per site $C/kN$ (lower panel) as a
function of temperature $1/J$ at fixed antiferromagnetic bond
concentration $p=0.5$, for $d=3$ systems with underfrustration
$(f=0.02,0.2,0.5)$, the stochastic frustration $(f=1=g)$, and
overfrustration $(g=0.7)$. The tick mark shows the phase transition
point between the spin-glass phase and the paramagnetic phase for
each frustration case. It is seen that frustration lowers this
transition temperature. Thus, for stochastic frustration ($f=1=g$),
the specific heat peak occurs outside the spin-glass phase,
indicating that considerable short-range ordering occurs at higher
temperatures before the onset of spin-glass long-range order. By
contrast, for the more underfrustrated cases $(f=0.02,0.2)$, the
specific heat peak occurs inside the spin-glass phase, indicating
that considerable short-range disorder persists into the higher
temperatures of the spin-glass phase. This conclusion is also
reached from the entropy curves in the upper panel. The changeover
between these two regimes occurs at the underfrustrated system of
$f=0.5$. Overfrustrated systems show understandably specific heat
behavior similar to $f=1$, with frustration lowering the long-range
order temperature and short-range order setting at higher
temperatures with a specific heat peak.}
\end{figure}

\begin{figure}[h!]
\centering
\includegraphics[scale=1]{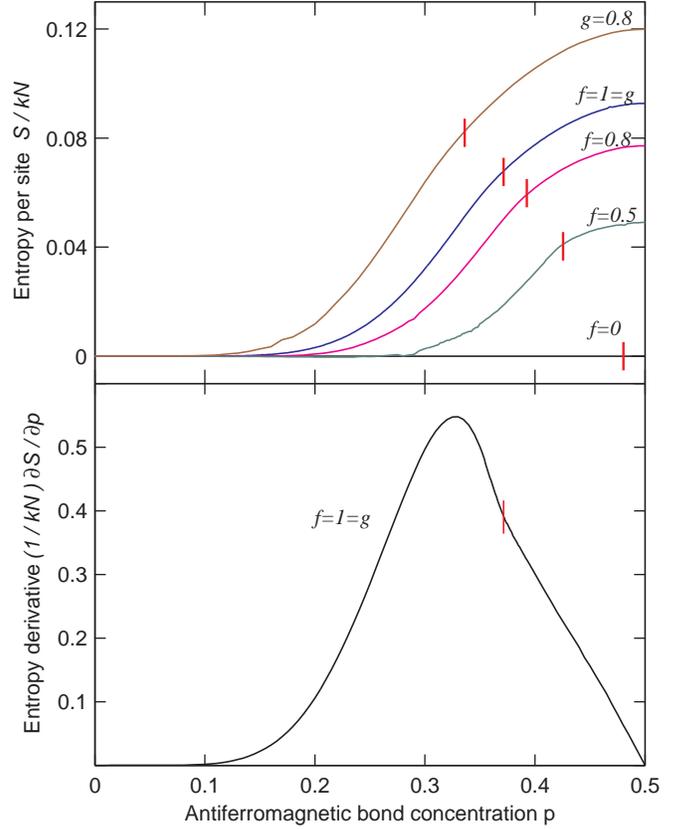}
\caption{(Color online) Top panel: The calculated entropy per site
$S/kN$ as a function of the antiferromagnetic bond concentration $p$
at fixed temperature $1/J=0.5$, for systems with no frustration
$(f=0)$, underfrustration $(f=0.5,0.8)$, the stochastic frustration
$(f=1=g)$, and overfrustration $(g=0.8)$. The tick mark shows the
phase transition point between the ferromagnetic phase and the
spin-glass phase for each frustration case. It is seen that
frustration favors the spin-glass phase over the ferromagnetic
phase.  It is also seen that, as soon as frustration is introduced,
the major portion of the entropy is created with the ferromagnetic
phase as opposed to the spin-glass phase. Lower panel: The
calculated derivative of the entropy per site $(1/kN)(\partial S /
\partial p)$ as a function of the antiferromagnetic bond
concentration $p$ at temperature $1/J=0.5$, for the stochastic
frustration system $(f=1)$ in $d=3$. The tick mark shows the phase
transition point between the ferromagnetic phase and the spin-glass
phase.  The peak being inside the ferromagnetic phase shows that
short-range disorder sets inside the ferromagnetic phase.}
\end{figure}

\section{Conclusion}

This study has started upon the realization that in Ising spin
glasses, frustration can be adjusted continuously and, if needed,
considerably, without changing the antiferromagnetic bond
probability $p$, by using locally correlated quenched randomness, as
we demonstrated here on hypercubic lattices and hierarchical
lattices. Thus, a rich variety of new spin-glass models and
spin-glass phases was created.  Such overfrustrated and
underfrustrated systems on hierarchical lattices in $d=3$ and 2 were
studied in detail, yielding new information and insights.  With the
removal of just 51\% of frustration $(f=0.49)$, a spin-glass phase
appears in $d=2$. With the addition of just 33\% frustration
$(g=0.67)$, the spin-glass phase disappears in $d=3$.  Sequences of
phase diagrams for different levels of frustration have been
calculated in both dimensions.  In general, frustration lowers the
spin-glass ordering temperature.  At low temperatures, frustration
favors the spin-glass phase (before it disappears) over the
ferromagnetic phase and symmetrically the antiferromagnetic phase.

When any amount, including infinitesimal, frustration is introduced,
the chaotic rescaling of local interactions occurs in the spin-glass
phase.  Chaos increases with increasing frustration, as seen from
the increased positive value of the calculated Lyapunov exponent,
starting from zero when frustration is absent.  The calculated
runaway exponent of the renormalization-group flows decreases, from
$y_R=d-1$ with increasing frustration to $y_R=0$ when the spin-glass
phase disappears.

From our calculations of entropy and specific heat curves in $d=3$,
it is seen that frustration lowers in temperature the onset of both
long- and short-range order in spin-glass phases, but is more
effective on the former.  Thus, for highly overfrustrated cases,
considerable short-range order occurs in the lower temperature range
of the paramagnetic phase, whereas for moderately overfrustrated,
stochastically frustrated, and underfrustrated cases, considerable
short-range disorder occurs in the higher temperature of the
spin-glass phase.  From calculations of the entropy and its
derivative as a function of antiferromagnetic bond concentration
$p$, it is seen that the ground-state and low-temperature entropy
already mostly sets in within the ferromagnetic and
antiferromagnetic phases, before the spin-glass phase is reached.

It is hoped that these calculational results, strictly valid for
hierarchical lattices but suggestive for hypercubic lattices, would
be repeated by Monte Carlo simulation, or other methods, for
hypercubic lattices, as we have demonstrated the preparation of
overfrustrated and underfrustrated hypercubic lattices.

\begin{acknowledgments}
Support by the Alexander von Humboldt Foundation, the Scientific and
Technological Research Council of Turkey (T\"UBITAK), and the
Academy of Sciences of Turkey (T\"UBA) is gratefully acknowledged.
\end{acknowledgments}

\end{document}